# Origin of the enhanced flexoelectricity of relaxor ferroelectrics

**Authors**

Jackeline Narvaez[1,2,+], Gustau Catalan[1,3,*]

[1] ICN2 – Institut Catala de Nanociencia i Nanotecnologia, Campus UAB, 08193 Bellaterra (Barcelona), Spain.

[2] CSIC - Consejo Superior de Investigaciones Cientificas, ICN2 Building, Campus UAB, 08193 Bellaterra (Barcelona), Spain.

[3] ICREA - Institucio Catalana de Recerca i Estudis Avançats, 08010 Barcelona, Spain.

[+]jackeline.narvaez@cin2.es

[*] gustau.catalan@cin2.es

**Abstract**

We have measured the bending-induced polarization of $Pb(Mg_{1/3}Nb_{2/3})O_3$-$PbTiO_3$ single crystals with compositions at the relaxor-ferroelectric phase boundary. The crystals display very large flexoelectricity, with flexocoupling coefficients an order of magnitude bigger than the theoretical upper limit set by the theories of Kogan and Tagantsev. This enhancement persists in the paraphrase up to a temperature T* that coincides with the start of elastic softening in the crystals. Analysis of the temperature dependence and cross-correlation between flexoelectric, dielectric and elastic properties indicates that the large bending-induced polarization of relaxor ferroelectrics is not caused by intrinsically giant flexoelectricity, but by the reorientation of polar nanotwins that become ferroelastically active below T*.

The giant electromechanical performance of relaxor-based ferroelectrics [1] and the complex physics associated with their inherently nanoscopic phase separation have inspired much research into these compounds [2]. The archetypal relaxor, Pb(Mg$_{1/3}$Nb$_{2/3}$)O$_3$ (PMN), was also the first ceramic for which bending-induced polarization (flexoelectricity) [3, 4] was measured [5], and it was the unexpectedly large value of the flexoelectric coefficient in Pb(Mg$_{1/3}$Nb$_{2/3}$)O$_3$ that triggered the investigation of flexoelectricity in other perovskite ferroelectrics such as Pb(Zr,Ti)O$_3$ [6,7], BaTiO$_3$[8] and (Ba,Sr)TiO$_3$[9]; these investigations, combined with the realization that very large flexoelectric effects can be achieved in the nanoscale [10, 11, 12] have contributed to the current surge of interest in this phenomenon.

Yet, for all the research, we still do not know the actual intrinsic value of the effective flexoelectric coefficients –the constants of proportionality between strain gradient and induced polarization. With the exception of SrTiO$_3$[13], for most perovskites the experimentally measured flexoelectricity exceeds theoretical expectations by between one and three orders of magnitude [14, 15, 16, 17], and the differences are not merely between theory and experiment, as experimental results can also substantially disagree among themselves; in Pb(Mg$_{1/3}$Nb$_{2/3}$)O$_3$-10%PbTiO$_3$, for example, there is a discrepancy of three orders of magnitude between flexoelectric coefficients measured by two different methods [18]. Meanwhile, the expected contribution of polar nanoregions to the flexoelectricity of relaxor ferroelectrics [3] has not been established, nor are there any measurements for compositions at or near the morphotropic phase boundary, even though their otherwise record-high electromechanical strain [1] might suggest the possibility of similarly enhanced flexoelectric effects. In addition, most flexoelectric measurements to date have been performed in ceramics, and there are no experimental reports for any single crystals other than SrTiO$_3$[13, 19]; this is relevant because grain boundaries have their own piezoelectric properties [20, 21] that can add an extrinsic contribution to the bending-induced polarization.

In this context, we have studied the bending-induced polarization of relaxor-ferroelectric single crystals with compositions (1-x)Pb(Mg$_{1/3}$Nb$_{2/3}$)O$_3$-xPbTiO$_3$, with x=0.28 and 0.34 (hereafter labelled PMN-28%PT and PMN-34%PT). These crystals, commercially available from TRS [22], are at the morphotropic phase boundary that separates a relaxor-like rhombohedral phase for PMN-rich compositions from a ferroelectric tetragonal phase for PT-rich compositions[23]. We have found that the effective flexoelectric and flexocoupling coefficients are indeed very large: the flexoelectric coefficient is ten times bigger than that of pure PMN ceramics, and the flexocoupling voltage also exceeds theoretical expectations by an order of magnitude [24]. However, these enhanced coefficients are history-dependent, suggesting a role of domains. Analysis of the simultaneously measured elastic response confirms a direct correlation between the enhancement of flexoelectricity and the onset of mechanical softening in the materials, itself caused by the existence of nanodomains that are simultaneously ferroelectric and ferroelastic (polar nanotwins) up to a temperature T* that is well above that of the dielectric maximum [25,26]. The large bending-induced polarization is thus not due to intrinsic flexoelectricity, but to a bending-induced reorientation of pre-existing polar nanotwins.

Bending-induced polarization has been measured using the method described by Zubko *et al.*[13]: a dynamic mechanical analyzer (DMA 8000, Perkin-Elmer) is used to apply a periodic three-point bending stress between room temperature and 573K (300 Celsius), using a ramp-rate of 3 K/min, whilst simultaneously recording the elastic response (storage modulus and elastic loss).The DMA's mechanical force signal was fed into the reference channel of a lock-in amplifier (Stanford Research Instruments, model 830), while the samples' electrodes are connected to the measurement channel of the lock-in amplifier, which records the bending-induced displacement currents. The current is converted into polarization using $P = I/(2\pi\nu A)$, where $\nu$ is the frequency of the bending force and A is the area of the electrodes. The polarization measured by the lock-in is related to the effective flexoelectric coefficient $\mu_{13}^{eff}$ by

$$\bar{P}_3 = \mu_{13}^{eff} \overline{\frac{\partial \epsilon_{11}}{\partial x_3}} \quad \text{and} \quad \overline{\frac{\partial \epsilon_{11}}{\partial x_3}} = \frac{12 z_0}{L^3}(L - a)$$

Where *L* is the separation between the standing points of the crystal, *a* is the half-length of the electrodes, and $z_0$ is the maximum vertical deflection in the middle. Typical values in our measurements are L=8mm, a=3mm and $z_0$=2μm. The flexoelectric tensor components are always coupled together and cannot be individually measured in quasistatic bending experiments [19,27]; it is customary instead to define an effective coefficient that is a combination of the tensor components relating the strain gradient to the induced polarization. In the three-point bending geometry of our experiment, the effective flexoelectric coefficient is $\mu_{13}^{eff} \equiv \frac{P_3}{\partial s_1/\partial x_3}$, where sample length is along $x_1$, width along $x_2$ and thickness along $x_3$. For an isotropic material, $\mu_{13}^{eff}$ is related to the flexoelectric tensor components by [13, 24, 17] $\mu_{13}^{eff} = \mu_{11} + (1-\nu)\mu_{13}$, where $\nu$ is the Poisson's ratio.

Flexoelectricity is proportional to dielectric permittivity [3, 4, 24, 28], so it is useful to also characterize the dielectric constant of the crystals. The dielectric constant and loss as a function of temperature (Figure 1) were measured using an Agilent E4980A Precision LCR-meter to measure the capacitance and the DMA receptacle to control the temperature. The dielectric losses are low (tan$\delta$<0.05) for both samples throughout the entire temperature range of the experiments; this ensures that the impedance response is predominantly dielectric even at the highest recorded temperatures, where dielectric losses begin to rise due to increased conductivity. The samples display clear dielectric maxima, but with some differences: the peak of PMN-34%PT is sharper and at higher temperature than that of PMN-28%PT. This is to be expected and correlates with the concentration of PbTiO$_3$ (PT), which is a standard ferroelectric with a high Curie temperature ($T_C$=492$^o$C) [29]. The PMN-34%PT sample has a dielectric response closer to that of ferroelectric PbTiO$_3$ (sharp peak, higher temperature Tc=150$^o$C) while the PMN-28%PT sample has response closer to the relaxor Pb(Mn$_{1/2}$Nb$_{2/3}$)O$_3$, with concomitantly lower peak temperature (Tm=125$^o$C) and increased diffuseness. Despite the difference in diffuseness, the inverse permittivity (inset of figure 1) shows a clear departure from linear Curie-Weiss behaviour for both compositions. Such departure is a classic indicator of the existence of polar nanoregions, characteristic of relaxors, in the temperature range above the dielectric maximum and up to a temperature between 225-250 Celsius [30]. Thus, irrespective of whether the dielectric peak is diffuse (canonical relaxor) or sharp (ferroelectric), the high temperature phase above the peak appears to be the same for both PMN-28%PT and PMN-34%PT.

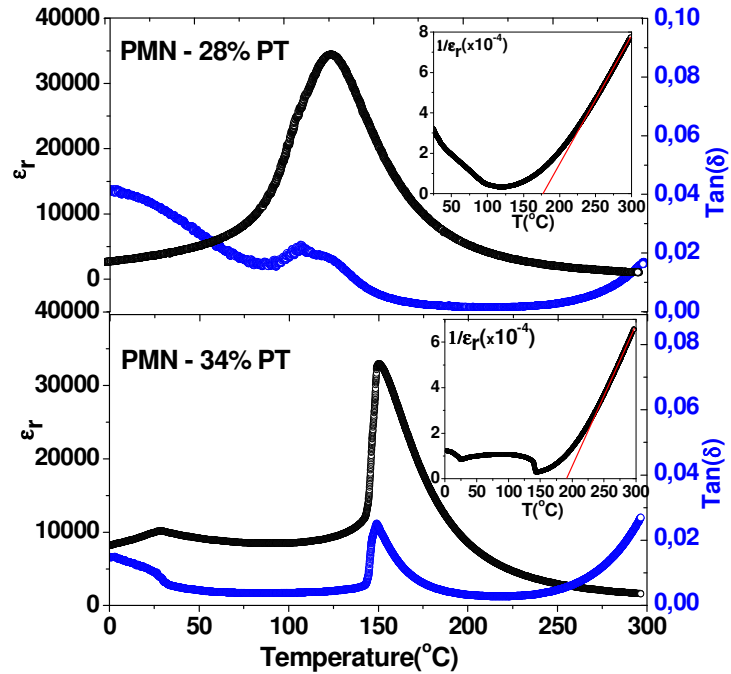

**Figure 1.** Dielectric constant and loss of PMN-28%PT and PMN-34%PT. Insets: inverse of the relative dielectric constant, showing a departure from linear Curie-Weiss behavior.

The effective flexoelectric coefficients as a function of temperature are shown in Figure 2. The Young's modulus and elastic loss tangent, measured simultaneously with the flexoelectric coefficient, are also included. It can readily be seen from the temperature dependences in Figure 2 that the increase of the flexoelectric coefficient is correlated with a decrease of the elastic constant. The flexoelectric coefficients display maxima at the same temperatures as the dielectric permittivity, with maximum values around 30μC/m, to be compared with c.a. 12μC/m for PMN-10%PT [18] and 2μC/m for PMN[5]; thus, compositional closeness to the morphotropic phase boundary indeed leads to bigger effective flexoelectric coefficients, although these values are still below those of the best known flexoelectric material, (Ba,Sr)TiO$_3$ [9,24,31]. Importantly, though, we also note that the large flexoelectric coefficients are dependent on the thermal history of the sample: up to a temperature T*~225±25°C, flexoelectricity is higher on heating than on cooling, which suggests a role of domains, even in the nominally paraelectric phase, up to T*.

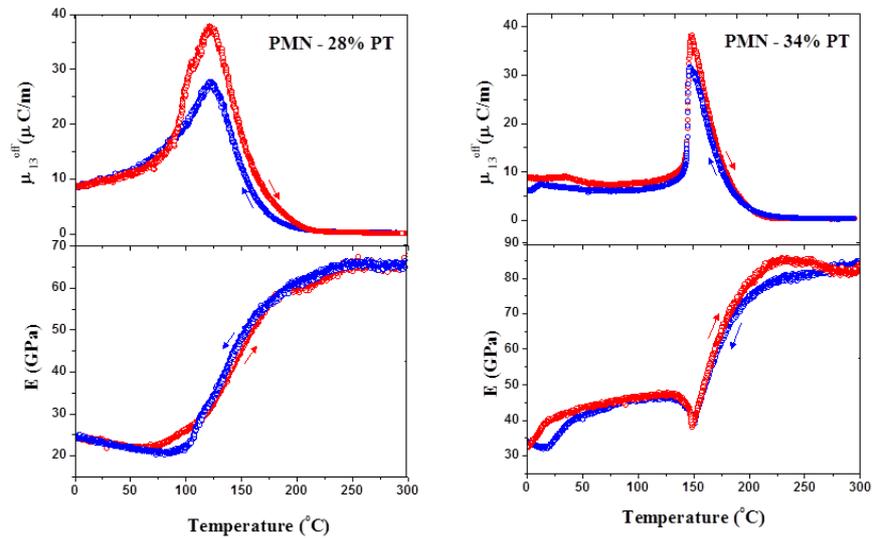

**Figure 2.** Temperature dependence of the flexoelectric coefficients and elastic Young's modulus of PMN-28%PT and PMN-34%PT.

Proportionality between flexoelectricity and permittivity can account for the coincidence between the peaks for dielecric constant and flexoelectricity. Thus, in order to disentangle the effect of permittivity and gain further insight into the high values of flexoelectricity, it is useful to examine the flexoelectric coefficient normalized by the dielectric constant. Theoretically, this flexo-coupling (or flexo-voltage[12]) coefficient $f$ should be of the order of 1-10V [24]. In our experiments (figure 3), the flexocoupling coefficients instead reach up to $f$~300 volts, which is an order of magnitude higher than the theoretical upper limit. Since these values are reached at or below the temperature of the dielectric maximum, it seems likely that they are related to ferroelectricity–even for PMN-28%PT, which is a relaxor, long range polar order can be field-induced below the freezing temperature. At temperatures above $T^*$, however, the flexocoupling coefficient decreases to a value $f \leq 10V$ that is not hysteretic and is consistent with theoretical expectations. This is therefore likely to be the true intrinsic value of the flexocoupling coefficient, with the enhancement below $T^*$ arising from additional contributions beyond pure flexoelectricity.

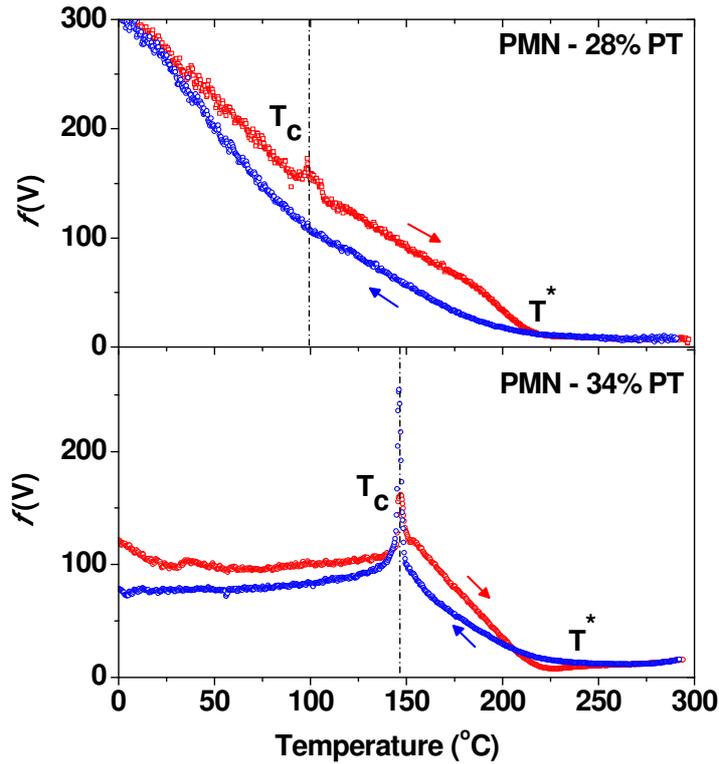

**Figure 3.** Flexocoupling coefficients of PMN-28%PT and PMN-34%PT

In order to understand this flexoelectric response, it is useful to re-examine the elastic behaviour in Figure 2. At high temperatures, the Young's modulus $E$ is relatively constant, but around 200-250°C the lattice begins to soften, with the Young's modulus decreasing from 60 GPa above T* to 20 GPa below Tm for PMN-28%PT, and from 80 GPa to 40 GPa for PMN-34%PT. Relaxor ferroelectrics are known to display a peak in acoustic emission at a relatively composition-independent temperature T*≈500±30K [25,26]; since acoustic emission is caused by a sudden release of elastic energy, T* must have a signature in the elastic constants, in agreement with our results, and consistent also with analysis by resonant ultrasound spectroscopy[34]. The origin of the peak in acoustic emission, and of the mechanical softening below T*, is a transition within the relaxor polar nanorregions whereby they become ferroelastically active at T*[25, 26, 35, 34].

The appearance of polar nanotwins below T* has profound consequences for the electromechanical response of the material, because now external stress can cause a ferroelastic reorientation of these domains and thus a local change of polarization. The mechanical strains induced by the bending reach up to $\epsilon_{11} \approx 10^{-4}$ in the surfaces. Multiplied by the Young's modulus of PMN-PT (20-80 GPa), this is equivalent to a stress of 2-8 MPa, which is sufficient to cause ferroelastic switching [36] and, combined with an electric bias, it can pole PMN-PT even at room temperature [36]. We emphasize, however, that an electric bias is still needed in order to remove the degeneracy between the +z and –z directions [37]. Stress alone can rearrange ferroelastic domains, but it cannot yield a net increase in polarization, because strain couples to the square of polarization and thus it cannot favour one polarity over its opposite [12, 24]. Under bending stress, local compression of the x-y plane at the concave side of the crystals must increase the proportion of domains with out-of-plane polarization, but in the absence of a bias the positive and negative domains would cancel each other and yield no net gain –this is the reason why mechanical poling is always done in the presence of an electric bias [36].

In the absence of electrostatic or built-in biases, then, the only other source of bias is the strain gradient. The bending-induced vertical strain gradient breaks the symmetry between upward and downward directions, favouring the polarity parallel to the flexoelectric field. The strain gradient in our samples is of the order of $\frac{\partial \epsilon_{11}}{\partial z} \approx 0.1 \text{m}^{-1}$, and the intrinsic (high temperature) flexocoupling coefficient is of the order of f≈10V, so the equivalent flexoelectric field is of the order of $E = f \frac{\partial \epsilon_{11}}{\partial z} \approx 1$V/m. By itself, this flexoelectric field is too small to cause ferroelectric switching (typical poling fields in ferroelectrics are of the order of KV/cm), but it is sufficient to break the inversion symmetry of the crystal, so that, as the stress causes ferroelastic switching, the strain gradient loads the dice in the poling direction. Further evidence for the role of nanodomains on the enhanced flexoelectricity is also provided by the thermal history dependence of the flexocoupling coefficient between $T_C$ and T*: the coupling is always higher on heating than on cooling, reflecting the existence of a bigger volume fraction of nanodomains when heating from the ferroelectric polar state than when cooling from the non-polar paraphase. Though the potential usefulness of the enhanced flexoelectricity for electromechanical transduction is mostly unaffected by this mixed origin, the small dependence on thermal history, inherent in any domain-based property, is undesirable. Perhaps more importantly, the results show that giant bending-induced polarization may be induced without an actual giant flexoelectricity, and this should help reconcile present discrepancies about the true size of the flexoelectric coefficient.

In summary: The flexoelectricity of PMN-PT can be very big –the flexoelectric coefficient is ten times larger than those of pure PMN, and the flexocoupling coefficient is also an order of magnitude higher than the upper limit set by the models of Kogan and Tagantsev [24]. This enhanced flexocoupling, however, appears only at temperatures below T*≈250±25°C and seems to be due to bending-induced switching of polar nanodomains that become feroelastically active around T*. "Pure" flexoelectricity (lattice-based as opposed to domain-based) can be measured above T*, yielding stable flexocoupling coefficients around $f_{13}$≈10V for both compositions. These values are comparable to those of other perovskite ferroelectrics and fully in line with theoretical expectations [24]. For the flexoelectric enhancement below T*, ferroelasticity and flexoelectricity must work in tandem; stress alone can cause domain switching, but no net change in polarization, while the flexoelectric field is not strong enough to pole the sample by itself. It is the combined effect of ferroelastic switching within a flexoelectric bias that causes the net change in polarization, (figure 3). Thus, while the effective flexoelectric coefficients may be very large, the intrinsic ones are not: the enhanced performance is caused by the coupling of flexoelectricity to the nanoscopic ferroelectricity inherent in relaxors.

## Acknowledgements

This work was funded by an ERC Starting Grant from the EU (project "Flexoelectricity"), and by Project MAT2010-17771from the Spanish Ministry of Education. We are grateful to Dr Neus Domingo for useful discussions in the preparation of this manuscript.